# Carrier Lifetime Enhancement in a Tellurium Nanowire/PEDOT:PSS Nanocomposite by Sulfur Passivation

James N. Heyman[1], Ayaskanta Sahu[2], Nelson E. Coates[2], Brittany Ehmann[1], Jeffery J. Urban[2]

1. Physics Department, Macalester College, St. Paul, MN 55105, USA
2. The Molecular Foundry, Materials Science Division, Lawrence Berkeley National Laboratory, Berkeley, CA, 94720, USA

**ABSTRACT**

We report static and time-resolved terahertz (THz) conductivity measurements of a high-performance thermoelectric material containing tellurium nanowires in a PEDOT:PSS matrix. Composites were made with and without sulfur passivation of the nanowires surfaces. The material with sulfur linkers (TeNW/PD-S) is less conductive but has a longer carrier lifetime than the formulation without (TeNW/PD). We find real conductivities at $f = 1$ THz of $\sigma_{TeNW/PD} = 160$ S/cm and $\sigma_{TeNW/PD-S} = 5.1$ S/cm. These values are much larger than the corresponding DC conductivities, suggesting DC conductivity is limited by structural defects. The free-carrier lifetime in the nanowires is controlled by recombination and trapping at the nanowire surfaces. We find surface recombination velocities in bare tellurium nanowires (22m/s) and TeNW/PD-S (40m/s) that are comparable to evaporated tellurium thin films. The surface recombination velocity in TeNW/PD (509m/s) is much larger, indicating a higher interface trap density.

**INTRODUCTION**

Improvements in the efficiency of thermoelectric materials would open the door to a wide range of significant applications in energy production, waste energy capture and thermal management. The efficiency of a thermoelectric generator or refrigerator is controlled by the thermoelectric figure of merit of its constituents, $ZT = S^2 \sigma T / \kappa$. Nanocomposite materials created by embedding nanoparticles in a suitable matrix have attracted great interest as possible thermoelectric materials[1-3]. In these materials, a high density of interfaces is available to scatter phonons so as to effectively suppress $\kappa$. Nanocomposites also offer many material parameters that can be adjusted to optimize the thermoelectric power factor $S^2 \sigma$. From a practical standpoint, many nanocomposites can be fabricated by solution processing and other techniques that are inherently inexpensive and scalable. However, efficient thermoelectric materials must also be good conductors. It is critical to engineer efficient carrier transport in nanocomposite materials by building them from conductive components and suppressing electron scattering at nanoparticle/matrix interfaces.

We investigated two nanocomposite materials consisting of tellurium nanowires coated with PEDOT:PSS. The first material (TeNW/PD) was produced by a one-pot synthesis based on a previously established nanowire synthesis with the substitution of PEDOT:PSS as the structure-directing agent [4-6]. Our PEDOT:PSS (PD) was purchased from Clevios™ and is the PH1000 formulation. Samples investigated in this work had nanowire diameters of ~12nm and a tellurium volume fraction of approximately 50%. The second material (TeNW/PD-S) was produced in two steps: first, bare Te nanowires were synthesized following established synthetic protocols (doi: 10.1002/1521-4095(200109)); once synthesized, the nanowire surfaces were functionalized with sulfur ($S_2^-$) ions by mixing a dispersion of nanowires with sodium sulfide ($Na_2S$) in water for a few minutes. After cleaning out excess $Na_2S$ by repeated centrifugation, PEDOT:PSS was added to the Te-($S_2^-$) nanowires and the mixture was allowed to stir overnight. Excess unbound PEDOT:PSS was again removed by repeated centrifugation and re-dispersion in water. These samples had nanowire diameters of ~35nm, and a tellurium volume fraction of 80-85%. Bare nanowires (TeNW-S) produced by the second procedure and functionalized with sulfur ions were also investigated.

We use THz spectroscopy to characterize the frequency-dependent conductivity of these materials. While DC conductivity ($\sigma_{DC}$) is most relevant to thermoelectric performance, $\sigma_{DC}$ is strongly dependent on a sample's large-scale structure and morphology, which can mask the microscopic transport phenomena. THz spectroscopy[7] offers an ultra-high frequency, contact-free conductivity probe that is less sensitive to large-scale defects (cracks, etc.) and film morphology (grain size, linkage between grains) than $\sigma_{DC}$. The reasons for this are straightforward: large scale features such as scratches and cracks do not greatly perturb optical measurements if their density is low. In addition, the size and shape of sub-wavelength sized grains is unimportant in AC conductivity measurements at frequencies that exceed the dielectric relaxation rate $f >> \sigma/(2\pi\varepsilon)$.

We have also used Optical Pump/THz probe time-resolved transmission measurements to study carrier dynamics in our samples. In this technique an optical pump-pulse excites free carriers in a material and the transmission of a delayed THz probe-pulse is used to monitor the evolution of the conductivity. These measurements directly determine carrier cooling and recombination rates.

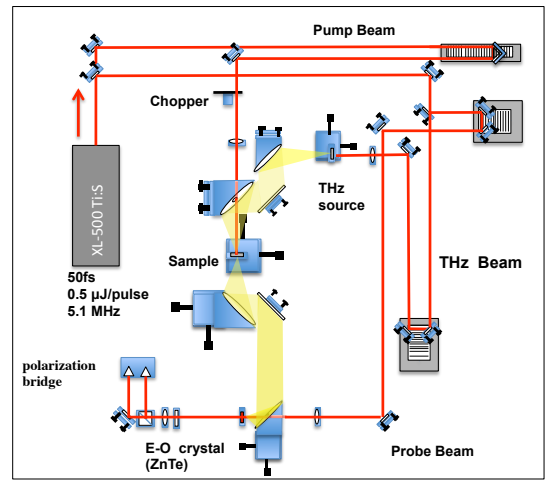

**Figure 1.** Layout of the Optical Pump/THz Probe experiment. The THz transmission experiment is similar but uses a 10fs pulse Ti:Sapphire oscillator and omits the pump beam.

**EXPERIMENTAL DETAILS**

THz transmission measurements were performed with a time-domain THz system based on a Ti:Sapphire oscillator ($\tau$=10fs, $\lambda$=800nm, 75MHz repetition rate) (Fig. 1). The THz emitter was a photoconductive switch. A 0.3 mm-thick GaP crystal was used as an electro-optic THz detector. The spectrometer was purged with dry air to suppress THz absorption by water vapor. We measured the electric field versus time of THz pulses transmitted through the samples and through a bare silicon or quartz reference sample. The Fourier Transform of each waveform yields a single-beam spectrum (amplitude and phase). The complex transmission is the ratio of sample and reference spectra. The frequency-dependent conductivity was extracted from the complex transmission by considering the sample as a thick conducting film on a dielectric substrate [8].

Figure 2 shows the complex conductivity versus frequency for our samples. The real conductivity is approximately constant between $f$ = 1 to 2 THz in both materials, with $\sigma'_{TeNW/PD}$ = 160 S/cm and $\sigma'_{TeNW/PD-S}$ = 5.1 S/cm at $f$ = 1THz. Both materials show a suppression of the conductivity at low frequencies. The strong feature observed at 2.7THz in the (thicker) TeNW/PD-S sample is an infrared-active tellurium phonon. Measurement error is dominated by uncertainty due to variations in sample thickness.

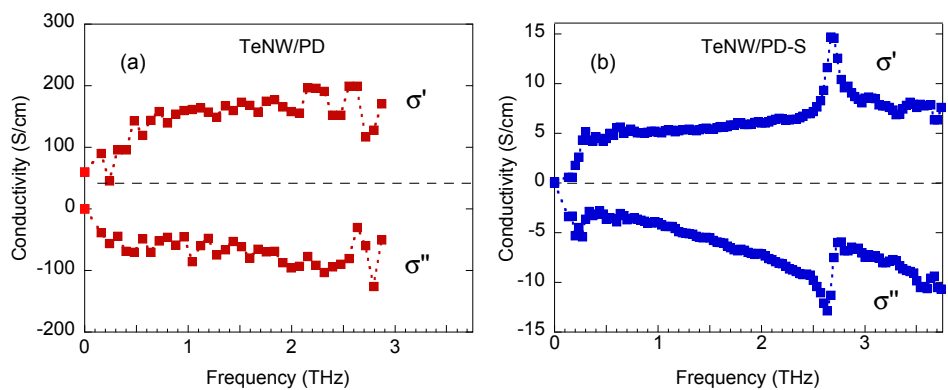

**Figure 2.** Complex conductivity ($\tilde{\sigma} = \sigma' + i\sigma''$) vs. frequency for (a) TeNW/PD and (b) TeNW/PD-S

Our time-resolved THz system is based on a FemtoLasers XL500 chirped-pulse oscillator (800nm center wavelength, 5MHz repetition rate, 0.5μJ pulse energy, 50fs pulse-width) (Fig. 1). The laser pulses are divided into Pump, THz Generation and Probe beams, with each beam incorporating an independent delay stage. THz probe pulses are generated and detected as in the transmission experiment above. The pump pulse excites the samples at an incident photon flux up to $\phi \sim 4 \cdot 10^{13} cm^{-2}$ at 800nm. The pump beam is chopped and the signal is recovered using a lock-in amplifier. Sweeping the pump delay allows measurement of the change in THz transmission as a function of time following photoexcitation. For the small transmission changes observed here ($\Delta T$ < 10%) the pump-induced increase in conductivity is proportional to the decrease in transmission. We used a half-wave plate to rotate the pump polarization relative to the THz probe.

We performed optical pump/THz probe measurements on three samples for this study: the one-pot synthesis material TeNW/PD, the sulfur-linked material TeNW/PD-S, and a film of bare Te nanowires (TeNW-S) (Fig. 3). The dynamics following photoexcitation are similar in all three samples: the transmission decreases during the initial ~1 to 10ps and then recovers on a longer timescale. The pump-induced transmission change can be described by a simple expression,

$$\langle \Delta T \rangle = T_0 \left(1 - e^{-t/\tau_C}\right) e^{-t/\tau_R} \quad , \tag{1}$$

where $\tau_C$ is a carrier cooling time and $\tau_R$ is the conductivity decay time governed by photocarrier trapping and recombination. The photoconductivity decay time is strongly dependent on the synthesis method (Table 1). In the sulfur-linked material TeNW/PD-S and the bare nanowire sample TeNW-S the decay time is several hundred picoseconds. However, $\tau_R$ is strongly suppressed in the one-pot synthesis material TeNW/PD. The carrier cooling time $\tau_C$ is approximately 2 – 4ps in all samples investigated, much longer than the time resolution of our system (~0.2ps).

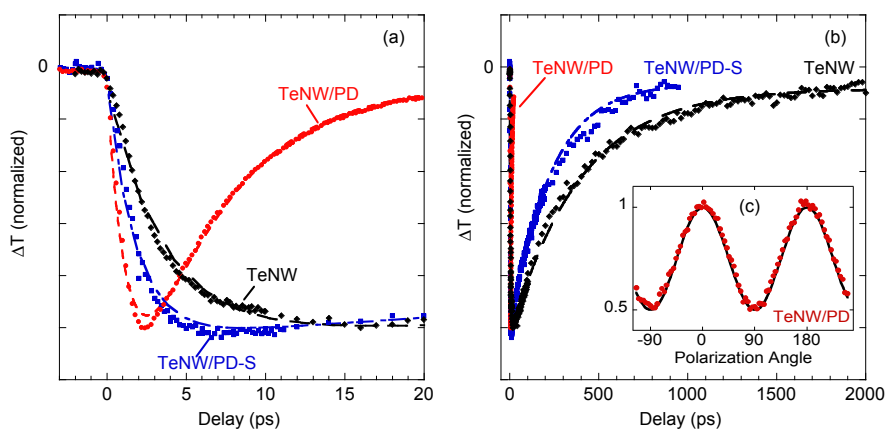

**Figure 3.** (a) Time-resolved THz transmission *vs.* delay after photoexcitation for delays <20ps, and (b) delays <2000ps. (c) Transmission change *vs.* polarization angle between pump and THz beams for TeNW/PD. Solid line is our model.

We also observed a strong dependence of the transient signal on the angle between the polarizations of the pump and THz beams. Basic electromagnetic theory[9] shows that isolated tellurium nanowires should primarily absorb pump radiation polarized parallel to the nanowire axis, and that photoexcitation should only strongly modify the THz conductivity parallel to the nanowire axis. In this case the variation should fit the expression $\langle |\Delta T| \rangle \propto \frac{1}{2}(1 + \cos^2 \theta)$. The excellent fit between this simple model and the data shows that the time-resolved signal is dominated by photoconductivity in the nanowires.

**DISCUSSION**

The principle result of the time-resolved THz transmission measurements is that the photoconductivity decay time is much longer in TeNW/PD-S compared to TeNW/PD. Our results show that sulfur functionalization reduces the density of traps and recombination centers at the nanowire surfaces.

Strong indirect evidence suggests that the photoconductivity decay time is controlled by recombination and trapping at nanowire surfaces: we observe longer lifetimes (>1ns) in annealed material containing micron-diameter Te nanowires; we observe a ~50% reduction in lifetime in oxidized *vs.* pristine nanowires; and lifetimes are much shorter than the ~10μs carrier lifetime reported [10] in high quality bulk Te.

In this case we can extract a geometry-independent surface recombination velocity $v_s$ from the photoconductivity decay times, $v_s = R/2\tau_R$ where R is the wire radius and we have assumed that carrier diffusion across our cylindrical nanowires is much faster than the photoconductivity decay. An Einstein-model of the diffusion coefficient supports this assumption for carrier mobility $\mu > 10 cm^2/Vs$. The recombination velocities we obtain (Table 1) for bare TeNW-S and TeNW/PD-S are comparable to the reported[11] value for evaporated high-purity Te thin-films of $v_S = 60$m/s. In contrast, $v_S$ for TeNW/PD is more than an order of magnitude larger, indicating a higher density of traps and recombination centers at nanowire surfaces in this material.

Table I. Conductivity at $f = 1$THz, carrier cooling time $\tau_C$, photoconductivity decay time $\tau_R$ and recombination velocity for the samples investigated in this study.

| Sample | $\sigma'_{1THz}$ (S/cm) | $\tau_C$ (ps) | $\tau_R$ (ps) | $v_s$ (m/s) |
|---|---|---|---|---|
| TeNW/PD | 160 | 1.9 | 5.4 | 509 |
| TeNW/PD-S | 5.1 | 1.8 | 219 | 40 |
| TeNW-S | - | 3.6 | 390 | 22 |

## CONCLUSIONS

We have used THz frequency conductivity and time-resolved THz transmission measurements to study a nanocomposite thermoelectric material consisting of tellurium nanowires in a PEDOT:PSS matrix. Two formulations of this material were investigated: a one-pot synthesis (TeNW/PD), and a synthesis with sulfur-ion linkers bonding the nanowires to the PEDOT (TeNW/PD-S). TeNW/PD is found to be highly conductive (160S/cm) while TeNW/PD-S exhibits lower conductivity (5.1 S/cm). Both show a strong suppression of conductivity at low frequencies, which is likely due to structural imperfections or grain boundaries. We find that processes at the nanowire surfaces control the carrier lifetime in the nanowires. Long carrier lifetimes in sulfur functionalized bare nanowires (TeNW-S) and in TeNW/PD-S yield surface recombination velocities comparable to high-purity evaporated Te thin films. The recombination velocity in TeNW/PD an order of magnitude higher, implying a high density of recombination centers and traps at the nanowire surfaces.

## ACKNOWLEDGEMENTS

This research was supported by the National Science Foundation under Grant No. DMR-1006065 and by the Department of Energy BES-LBL Thermoelectrics Program. Work at the

Molecular Foundry was supported by the Office of Science, Office of Basic Energy Sciences, of the U.S. Department of Energy under Contract No. DE-AC02-05CH11231.

**REFERENCES**

1. Z. G. Chen, G. Han, L. Yang, L. N. Cheng and J. Zou, *Prog. Nat. Sci-Mater.* **22** (6), 535-549 (2012).
2. P. Pichanusakorn and P. Bandaru, *Mat. Sci. Eng.* R **67** (2-4), 19-63 (2010).
3. G. J. Snyder and E. S. Toberer, *Nature Materials* **7** (2), 105-114 (2008).
4. N. E. Coates, S. K. Yee, B. McCulloch, K. C. See, A. Majumdar, R. A. Segalman and J. J. Urban, *Adv. Mater.* **25** (11), 1629-1633 (2013).
5. K. C. See, J. P. Feser, C. E. Chen, A. Majumdar, J. J. Urban and R. A. Segalman, *Nano Letters* **10** (11), 4664-4667 (2010).
6. S. K. Yee, N. E. Coates, A. Majumdar, J. J. Urban and R. A. Segalman, *Phys. Chem. Chem. Phys.* **15** (11), 4024-4032 (2013).
7. M. C. Beard, G. M. Turner and C. A. Schmuttenmaer, *J. Phys. Chem.* **106**, 7146-7159 (2002).
8. J. N. Heyman, B. A. Alebachew, Z. S. Kaminski, M. D. Nguyen, N. E. Coates and J. J. Urban, *Appl. Phys. Lett.* **104** (14) (2014).
9. C. F. Bohren and D. R. Huffman, *Absorption and scattering of light by small particles*. (Wiley, New York, 1983).
10. J. I. Pankove, *Optical Processes in Semiconductors*, 2nd ed. (Dover Publications, 2010), 162.
11. N. G. Shyamprasad, C. H. Champness and I. Shih, *Infrared Phys*. **21**, 45-52 (1980).